\documentclass[12pt,a4paper]{article}
\usepackage[utf8]{inputenc}
\usepackage{amsmath,amssymb}
\usepackage{graphicx}
\usepackage{hyperref}
\usepackage{authblk}
\usepackage{xfrac}
\usepackage{xcolor}

\title{Density screening effects in the NJL model: \\ Chiral condensate, speed of sound, and high-density thermodynamics}
\author[1]{A. Rosas D\'iaz}
\author[2,3]{A. Raya}
\author[5,6,7]{C. A. Vaquera Araujo}
\author[2,4]{S. Hern\'andez-Ortiz}
\affil[1]{Instituto de F\'isica y Matem\'aticas, Universidad Michoacana de San Nicol\'as de Hidalgo, M\'exico}
\affil[2]{Facultad de Ingenier\'ia El\'ectrica, Universidad Michoacana de San Nicol\'as de Hidalgo, M\'exico}
\affil[3]{Centro de Ciencias Exactas, Universidad del B\'io-B\'io, Chile}
\affil[4]{Bogoliubov Laboratory of Theoretical Physics, JINR, Dubna, Russia}
\affil[5]{Secretar\'ia de Ciencia, Humanidades, Tecnolog\'ia e Innovaci\'on, Insurgentes Sur 1582, Colonia Cr\'edito Constructor, Benito Ju\'arez 03940, Ciudad de M\'exico, M\'exico}
\affil[6]{Departamento de F\'isica, Divisi\'on de Ciencias e Ingenier\'ias, Campus Le\'on, Universidad de Guanajuato, Loma del Bosque 103, Lomas del Campestre, Le\'on 37150, Guanajuato, M\'exico}
\affil[7]{Dual CP Institute of High Energy Physics, Colima 28045, Colima, M\'exico}

\date{November 2026}

\begin{document}
\maketitle

\begin{abstract}
The phase diagram of Quantum Chromodynamics (QCD) remains a central topic in high-energy physics. At high temperature and low baryochemical potential, the chiral transition is known to be a smooth crossover, while at high densities, a first-order phase transition is theoretically expected. In this work, we explore the dense, cold regime of the QCD phase diagram using the two-flavor Nambu--Jona-Lasinio (NJL) model, incorporating medium screening effects through a phenomenological density-dependent effective coupling . To isolate ultraviolet divergences without introducing 3D cutoffs that violate thermodynamics, we apply a modified Beta/Gamma-function regularization scheme. The in-medium contributions are evaluated using zero-temperature relations and full finite-temperature Fermi-Dirac integrals, incorporating the Hugenholtz--Van Hove rearrangement term dictated by the running coupling. Our numerical results indicate that the dynamically weakened interaction accelerates chiral symmetry restoration, shifting the transition toward lower quark chemical potentials. Conversely, the rearrangement term generates an effective repulsive contribution, significantly delaying the physical onset of quark matter and stiffening the equation of state in the transitional region. Consequently, when evaluated as a function of the baryon number density, the squared speed of sound exhibits a rapid rise at intermediate densities while respecting causality and smoothly asymptoting to the ultrarelativistic conformal bound. These findings provide a thermodynamically robust framework for cold dense quark matter, highlighting the crucial interplay between dynamically suppressed scalar interactions and emergent repulsive forces at high baryon densities.
\end{abstract}

\section{Introduction}

Understanding the phase structure of Quantum Chromodynamics (QCD) under extreme conditions
remains a central challenge in modern high-energy physics~\cite{Fukushima2011,BraunMunzinger2009}.
Lattice QCD calculations have established that, at vanishing baryochemical potential, the
transition from hadronic matter to the quark-gluon plasma (QGP) is a smooth
crossover~\cite{Aoki2006,Borsanyi2014,Bazavov2012}. The notorious sign problem, however,
prevents a reliable first-principles determination of the phase diagram at large baryon
density~\cite{deForcrand2009,Muroya2003}, leaving open fundamental questions about the
existence and location of a possible Critical End Point (CEP) and the order of the chiral
transition at high baryochemical potential~\cite{Stephanov2004}. Experimental programs such
as the Beam Energy Scan at RHIC and the upcoming measurements at FAIR and NICA aim to search
for signatures of the CEP~\cite{Luo2017,Adamczyk2014}, motivating parallel theoretical
studies using effective approaches that can access the high-density, low-temperature regime.

A variety of effective field-theoretical models and functional approaches have been developed
to map out the chiral and deconfinement structure of dense matter. Among the most widely used
are quark-meson (QM) and linear sigma models coupled to quarks, which treat scalar and
pseudoscalar mesons as explicit degrees of freedom coupled to constituent
quarks~\cite{Schaefer2007,Herbst2011,n97s-wqft}, as well as Polyakov-loop-extended frameworks, such as
the PNJL and PQM models, that incorporate an effective order parameter for deconfinement
alongside chiral dynamics~\cite{Fukushima2004,Ratti2006,Costa2008,fqkq-hs7h,knn8-sv3k}. Complementary continuum
approaches based on Dyson--Schwinger equations and the functional renormalization group further
bridge microscopic quark-gluon dynamics and non-perturbative phase
structure~\cite{Roberts1994,Fischer2009,Braun2012}. Within this landscape, effective models that
capture the chiral dynamics of QCD, notably the Nambu--Jona-Lasinio (NJL) model, remain
indispensable tools for exploring qualitative and semi-quantitative features of the QCD phase
diagram~\cite{Nambu1961,Klevansky1992,Hatsuda1994}.

The NJL framework successfully describes spontaneous chiral symmetry breaking and the associated
low-energy meson phenomenology, while being sufficiently tractable to allow controlled studies
of medium effects~\cite{Klevansky1992,Hatsuda1994,Vogl1991}. A common assumption in many NJL analyses is the use of a constant four-fermion coupling $G$~\cite{Buballa2005}. In a dense
medium, however, one expects the screening of interactions and an in-medium modification of
the effective coupling: neglecting such density-dependent screening may obscure important
structural modifications in the phase transition and the equation of state at high densities. In particular, ignoring medium screening can lead to an overestimation of the strength of chiral symmetry breaking and an underestimation of the stiffness of the equation of state.

In this work, we extend the two-flavor NJL model by incorporating medium screening effects through a density-dependent effective coupling $G'(\mu_q) = G\,m(\mu_q)/m_{\rm vac}$,
where $m_{\rm vac}$ is the dynamical mass in vacuum, $m(\mu_q)$ is the in-medium dynamical
mass, and $\mu_q = \mu_B/3$ is the quark chemical potential. This parametrization captures the
expected weakening of the scalar-channel attraction with increasing baryon density and is
physically motivated by the in-medium Debye screening of gluon exchanges. A microscopic
derivation of this coupling from Dyson-Schwinger equations or functional renormalization group
methods goes beyond the scope of the present paper, and the ansatz should be understood as a
minimal phenomenological model of in-medium screening; connections to related parametrizations
in the literature are discussed in Sec.~\ref{sec:discussion}. To ensure consistency across all
macroscopic variables and satisfy the Hugenholtz--Van Hove identity, the corresponding
rearrangement term is explicitly included in the evaluation of the baryon number density. Our
aim is to assess how this reduction of the scalar-channel attraction at large chemical potential
modifies the chiral condensate, the dynamical quark mass, and bulk thermodynamic observables
sensitive to the stiffness of the equation of state, in particular the squared speed of sound
$c_s^2$ as a function of the baryon density~\cite{Bedaque2015,Annala2020}.

We work in the Hartree approximation and solve the resulting gap equation,
$m = m_0 - 2G\langle\bar\psi\psi\rangle$. To bypass the unphysical artifacts associated with a
rigid three-momentum cutoff while retaining the quadratic momentum scale essential for
spontaneous chiral symmetry breaking, we apply a modified analytical Beta/Gamma-function scheme
characterized by an ultraviolet scale $\Lambda$ and an effective subtraction parameter
$\epsilon_{\text{eff}}$. While we present the Sommerfeld expansion in the theoretical framework
to provide analytic insight into the leading thermal and high-density corrections in the
degenerate regime ($\mu_q \gg T \sim 0$), our numerical calculations utilize the
zero-temperature analytical expressions and full finite-temperature Fermi-Dirac integrals,
avoiding asymptotic truncation.

First, our calculations show that density screening accelerates chiral symmetry restoration,
shifting the drop in the dynamical quark mass toward lower quark chemical potentials,
$\mu_q \approx 500$\,MeV. Second, the negative derivative of the running coupling
($\partial G'/\partial \mu_q < 0$) in the thermodynamic rearrangement term introduces an
effective repulsive contribution. This emergent repulsion delays the onset of quark matter from
$\mu_q \approx 301.4$\,MeV up to $\mu_q \approx 430$\,MeV and stiffens the equation of state.
Consequently, when mapped against the physical baryon number density $n_B$, the squared speed
of sound exhibits a steep rise at intermediate densities while respecting causality and smoothly
converging to the ultrarelativistic conformal bound ($c_s^2 \to 1/3$) at high densities.

The manuscript is organized as follows. Section~\ref{sec:framework} presents the theoretical
framework, details the modified analytical Beta/Gamma regularization scheme, defines the
dynamically screened coupling, and derives the thermodynamically consistent equation of state
including the rearrangement term. Section~\ref{sec:results} presents the numerical results for
the dynamical quark mass, the baryon number density, and the comparative analysis of the speed
of sound as a function of baryon density. Finally, Section~\ref{sec:discussion} discusses the
physical and astrophysical implications of our findings and outlines prospects for future work.

\section{Theoretical Framework}\label{sec:framework}

The starting point is the two-flavor Lagrangian for light quarks with scalar and
pseudoscalar interactions,
\begin{equation}
\mathcal{L}_{\rm NJL} = \bar{\psi}(i\gamma^{\mu}\partial_{\mu} - m_0)\psi
+ G\left[(\bar{\psi}\psi)^2 + (\bar{\psi} i\gamma_5 \vec{\tau}\psi)^2\right],
\label{eq:NJL}
\end{equation}
where $\psi$ is the quark field for $N_f = 2$ flavors and $N_c = 3$ colors, $m_0$ is the current quark
mass, $\vec{\tau}$ are the Pauli matrices in flavor space, and $G$ is the dimensionful
four-fermion coupling~\cite{Nambu1961}. Since the four-fermion interaction is non-renormalizable, an ultraviolet regularization prescription is required to render the loop integrals finite and introduce an effective physical scale $\Lambda$ ~\cite{Vogl1991,Klevansky1992,Hatsuda1994,Buballa2005}.

At low energies, the model generates a non-vanishing chiral condensate $\langle\bar\psi\psi\rangle\neq 0$,
reproducing the spontaneous breaking of chiral symmetry and providing a reliable description of
the light pseudoscalar meson sector~\cite{Klevansky1992,Vogl1991}. To evaluate the chiral condensate in the vacuum within the Hartree approximation~\cite{Hatsuda1994,Buballa2005}, we introduce the Feynman prescription $+i\varepsilon$ to define the contour integration:
\begin{equation}
    \langle \bar{\psi} \psi \rangle = -\mathrm{Tr}\int \frac{d^4k}{(2\pi)^{4}}\frac{i}{\not\! k-m+i\varepsilon} = -4iN_{f}N_{c}m \int \frac{d^{4}k}{(2\pi)^{4}}\frac{1}{k_{0}^{2} - \vec{k}^{2} - m^{2} + i\varepsilon}.
\end{equation}
Defining the on-shell energy $E_{k} = \sqrt{\vec{k}^{2} + m^{2}}$ and evaluating the contour integral over $k_0$ yields:
\begin{equation}
    \langle \bar{\psi} \psi \rangle = -2N_f N_c m \int \frac{d^3k}{(2\pi)^3} \frac{1}{E_k} = -\frac{N_f N_c m}{\pi^2} \int_0^\infty \frac{r^2}{\sqrt{r^2 + m^2}}\,dr.
\end{equation}
This integral contains both quadratic and logarithmic divergences. While pure dimensional or analytical regularization inherently discards quadratic divergences, the NJL model relies on this quadratic momentum scale to provide the necessary attractive strength for spontaneous chiral symmetry breaking; omitting it leads to unphysical inverse chiral catalysis at high densities.

We employ a modified Beta/Gamma-function subtraction scheme based on dimensional regularization~\cite{Klevansky1992,Fujihara2009,Inagaki2012,Kohyama2015}. By continuing the spatial dimension from $3$ to $3-2\epsilon$, the integration measure changes. To maintain the mass dimensions, we introduce a characteristic momentum scale $\Lambda$, which corresponds to the effective ultraviolet scale listed in Table~\ref{tab:model_parameters}, and modify the integration measure as $r^2 dr \to \Lambda^{2\epsilon} r^{2-2\epsilon} dr$, where the physical limit corresponds to $\epsilon \to 0$~\cite{Klevansky1992,Fujihara2009,Inagaki2012}:
\begin{equation}
    I_{\text{vac}} = \Lambda^{2\epsilon} \int_{0}^{\infty} \frac{r^{2-2\epsilon}}{\sqrt{r^{2} + m^{2}}}\,dr.
\end{equation}
Introducing the variable change $u = \sqrt{r^2+m^2}$, the integral becomes:
\begin{equation}
    I_{\text{vac}} = \Lambda^{2\epsilon} \int_{m}^{\infty} (u^2-m^2)^{\frac{1}{2}-\epsilon}\,du.
\end{equation}
Next, we use the substitution $v = 1 - \frac{m^2}{u^2}$ to obtain
\begin{equation}
    I_{\text{vac}} = \frac{m^{2-2\epsilon}}{2} \Lambda^{2\epsilon} \int_0^1 v^{\frac{1}{2}-\epsilon} (1-v)^{-2+\epsilon}\,dv.
\end{equation}
This form maps onto the Euler Beta function $\text{B}(x, y) = \frac{\Gamma(x)\Gamma(y)}{\Gamma(x+y)}$. Identifying $x = \frac{3}{2}-\epsilon$ and $y = -1+\epsilon$, we obtain the exact analytical result
\begin{equation}
    I_{\text{vac}} = \frac{m^{2-2\epsilon}}{2} \Lambda^{2\epsilon} \frac{\Gamma\left(\frac{3}{2}-\epsilon\right)\Gamma(-1+\epsilon)}{\Gamma\left(\frac{1}{2}\right)}.
\end{equation}
Around $\epsilon = 0$, we use $\Gamma(1/2) = \sqrt{\pi}$ and $\Gamma(3/2) = \frac{\sqrt{\pi}}{2}$. The simple pole at $z = -1$ yields the Laurent expansion $\Gamma(-1+\epsilon) = -\frac{1}{\epsilon} + \gamma_E - 1 + \mathcal{O}(\epsilon)$, where $\gamma_E \approx 0.5772$ is the Euler-Mascheroni constant. Expanding the scale factor gives $(\Lambda/m)^{2\epsilon} \approx 1 + 2\epsilon \ln(\Lambda/m)$. Thus, the divergent part of the integral reads:
\begin{equation}
    I_{\text{vac}} \approx -\frac{m^2}{4} \left[ \frac{1}{\epsilon} - \gamma_E + 1 - 2\ln\left(\frac{\Lambda}{m}\right) \right].
\end{equation}
As is standard in analytical continuation applied to non-renormalizable theories, pure dimensional regularization discards the quadratic divergence. To restore the phenomenological interaction scale required by the NJL model~\cite{Klevansky1992,Harada2003}, we manually reintroduce $\Lambda^2$ into the regularized vacuum integral. Furthermore, we absorb the dimensional pole $1/\epsilon$ and the accompanying constants into a single effective subtraction parameter defined as $\frac{1}{\epsilon_{\text{eff}}} = \frac{1}{\epsilon} - \gamma_E + 1 - 2\ln(\Lambda/m)$~\cite{Fujihara2009,Inagaki2012,Kohyama2015}. Thus, the regularized vacuum integral takes the form
\begin{equation}
    I_{\text{vac}}^{\text{reg}} = \Lambda^2 - \frac{m^2}{4} \left( \frac{1}{\epsilon_{\text{eff}}} \right).
\end{equation}
Substituting this back, the regularized vacuum condensate takes the form
\begin{equation}
    \langle \bar{\psi} \psi \rangle_{\text{vac}} = -\frac{N_f N_c m}{2\pi^2} \left[ \Lambda^2 - \frac{m^2}{4}\left(\frac{1}{\epsilon_{\text{eff}}}\right) \right].
\end{equation}
Using the gap equation $m = m_0 - 2G \langle \bar{\psi} \psi \rangle$, the vacuum constituent quark mass is generated by
\begin{equation}
   m_{\text{vac}} = m_0 + \frac{G N_{f}N_{c} m_{\text{vac}}}{\pi^{2}} \left[ \Lambda^2 - \frac{m_{\text{vac}}^2}{4} \left( \frac{1}{\epsilon_{\text{eff}}} \right) \right].
   \label{eq:gap_vac_reg}
\end{equation}
The effective subtraction parameter $\epsilon_{\text{eff}}$ is treated as a phenomenological constant, calibrated alongside the coupling $G$ and the scale $\Lambda$ to reproduce empirical vacuum observables: the pion mass $m_\pi$, the pion decay constant $f_\pi$, and the chiral condensate $\langle\bar{u}u\rangle^{1/3}$, yielding a vacuum constituent mass of $m_{\text{vac}} = 301.38$\,MeV. These values are listed in Table~\ref{tab:model_parameters}.

\begin{table}[htbp]
\centering
\caption{Parameter set of the two-flavor NJL model and the vacuum observables used to fix them. The current quark mass $m_0$, the effective ultraviolet scale $\Lambda$, the dimensionless coupling constant $G\Lambda^2$, and the effective subtraction parameter $1/\epsilon_{\text{eff}}$ reproduce the pion mass $m_\pi$, the pion decay constant $f_\pi$, and the chiral condensate in vacuum $\langle\bar{u}u\rangle^{1/3}$, generating the vacuum constituent mass $m_{\text{vac}}$.}
\label{tab:model_parameters}
\renewcommand{\arraystretch}{1.3} 
\begin{tabular}{lc @{\hspace{3em}} lc}
\hline\hline
\multicolumn{2}{l}{\textbf{Model Parameters}} & \multicolumn{2}{l}{\textbf{Vacuum Observables}} \\
\hline
$m_0$ & $5$\,MeV & $m_\pi$ & $138$\,MeV \\
$\Lambda$ & $1133.56$\,MeV & $f_\pi$ & $93$\,MeV \\
$G\Lambda^2$ & $2.428$ & $\langle\bar{u}u\rangle^{1/3}$ & $-250$\,MeV \\
$1/\epsilon_{\text{eff}}$ & $18.8863$ & $m_{\text{vac}}$ & $301.38$\,MeV \\
\hline\hline
\end{tabular}
\end{table}

At finite temperature $T$ and quark chemical potential $\mu_q = \mu_B / 3$, where $\mu_B$ is the baryonic chemical potential, medium effects are incorporated through the convergent Fermi-Dirac integral
\begin{equation}
    I_{\text{FD}}(m, \mu_q, T) = \int_{0}^{\infty} \frac{k^{2}}{\sqrt{k^{2} + m^{2}}} \left[n_{F}(k, \mu_q, T) + \bar{n}_{F}(k, \mu_q, T)\right] dk,
\end{equation}
where the occupation numbers are
\begin{equation}
    n_{F}(k, \mu_q, T) = \frac{1}{e^{(E_k - \mu_q)/T} + 1}, \qquad \bar{n}_{F}(k, \mu_q, T) = \frac{1}{e^{(E_k + \mu_q)/T} + 1}.
\end{equation}
For a degenerate medium with $\mu_q \gg m$ and finite $T \ll \mu_q$, antiparticles are exponentially suppressed ($\bar{n}_{F} \to 0$). Employing the Sommerfeld expansion~\cite{Ashcroft1976,FetterWalecka1971} up to $\mathcal{O}(T^2/\mu_q^2)$ for the convergent integral yields
\begin{equation}
\begin{split}
    I_{\text{FD}}(m, \mu_q, T)& \approx \int_{m}^{\mu_q} \sqrt{E^2 - m^2}\,dE + \frac{\pi^{2}T^{2}}{6} \frac{\mu_q}{\sqrt{\mu_q^2 - m^2}}\\&\equiv  I_0(m, \mu_q)+ \frac{\pi^{2}T^{2}}{6} \frac{\mu_q}{\sqrt{\mu_q^2 - m^2}}.
\end{split}
    \label{eq:sommerfeld}
\end{equation}
At zero temperature, the medium integral is given by the expression
\begin{equation}
    I_0(m, \mu_q) = \Theta(\mu_q - m) \left[ \frac{\mu_q}{2}\sqrt{\mu_q^2-m^2} - \frac{m^2}{2}\ln\!\left(\frac{\mu_q+\sqrt{\mu_q^2-m^2}}{m}\right) \right].
\end{equation}
In the high-density limit ($\mu_q \gg m$), $I_0$ simplifies to
\begin{equation}
    I_{0} \approx \frac{\mu_q^2}{2} - \frac{m^2}{4} - \frac{m^{2}}{2}\ln\!\left(\frac{2\mu_q}{m}\right),
\end{equation}
while the thermal correction simplifies to $\pi^2 T^2/6$. Thus, the asymptotic finite-density and temperature correction becomes
\begin{equation}
    I_{\text{FD}}(m, \mu_q, T) \approx \frac{\mu_q^2}{2} - \frac{m^2}{4} - \frac{m^{2}}{2}\ln\!\left(\frac{2\mu_q}{m}\right) + \frac{\pi^{2}T^{2}}{6}.
\end{equation}
Combining the regularized vacuum condensate with the medium contribution ($\langle \bar{\psi}\psi \rangle_{\text{med}} = \frac{N_f N_c m}{\pi^2} I_{\text{FD}}$), the full in-medium gap equation in the asymptotic regime reads
\begin{equation}\label{mFinal}
    m = m_0 + \frac{G N_{f}N_{c} m}{\pi^2} \left[ \Lambda^2 - \frac{m^2}{4}\left(\frac{1}{\epsilon_{\text{eff}}}\right) - \mu_q^2 + \frac{m^2}{2} + m^2\ln\!\left(\frac{2\mu_q}{m}\right) - \frac{\pi^{2}T^{2}}{3} \right].
\end{equation}

\subsection{Thermodynamic quantities and the screened coupling}

The thermodynamic potential in the Hartree approximation, extended to incorporate a medium-dependent effective interaction $G'(\mu_q)$, takes the standard form~\cite{Hatsuda1994,Buballa2005,Kapusta2006}
\begin{equation}
\Omega(T,\mu_q,m) = \Omega_{\rm vac}(m) + \Omega_{\rm med}(T,\mu_q,m) + \frac{(m-m_0)^2}{4G'(\mu_q)},
\end{equation}
where $\Omega_{\rm vac}$ collects the vacuum loop integral and $\Omega_{\rm med}$ the finite-density and thermal contributions. 

Using the modified analytical gap equation, the fermionic contributions to the potential are obtained by integrating the chiral condensate, $\langle \bar{\psi}\psi \rangle = -\partial \Omega_{\text{fermion}}/\partial m$. For the regularized vacuum part, integrating we obtain
\begin{equation}
\Omega_{\rm vac}(m) = \frac{N_f N_c}{2\pi^2} \left[ \frac{\Lambda^2 m^2}{2} - \frac{m^4}{16}\left(\frac{1}{\epsilon_{\text{eff}}}\right) \right].
\end{equation}
Similarly, the medium contribution is given by the exact Fermi-Dirac integral over the available phase space
\begin{equation}
\Omega_{\rm med}(T,\mu_q,m) = - 2 N_f N_c T \int \frac{d^3k}{(2\pi)^3} \left[ \ln\left(1 + e^{-(E_k - \mu_q)/T}\right) + \ln\left(1 + e^{-(E_k + \mu_q)/T}\right) \right].
\end{equation}

To maintain the standard form of the gap equation without introducing non-linear order-parameter derivatives, the screened coupling $G'$ is treated as an explicit function of the medium properties (the quark chemical potential $\mu_q$), evaluated along the physical trajectory
\begin{equation}
G'(\mu_q) \equiv G\,\frac{m(\mu_q)}{m_{\rm vac}},
\label{eq:Gprime}
\end{equation}
where $m(\mu_q)$ is the baseline solution of the gap equation at the given chemical potential and $m_{\rm vac}$ is the dynamical mass at $T=\mu_q=0$. This phenomenological parametrization interpolates between $G'=G$ in vacuum (where $m=m_{\rm vac}$) and $G'\to 0$ as chiral symmetry is restored and $m\to m_0$. It captures the qualitative expectation that the effective scalar interaction weakens in a dense medium, in analogy with the density dependence of the Debye screening mass in QCD~\cite{Roberts1994,Fischer2009}. 

Because the coupling $G'(\mu_q)$ runs with the chemical potential, a rearrangement term must be included for the Hugenholtz--Van Hove identity~\cite{Hugenholtz1958} to hold. The quark number density $n_q$ is given by the total derivative of the potential
\begin{equation}
n_q = -\frac{d\Omega}{d\mu_q} = -\left.\frac{\partial \Omega}{\partial \mu_q}\right|_{m, G'} - \left.\frac{\partial \Omega}{\partial m}\right|_{\mu_q, G'} \frac{\partial m}{\partial \mu_q} - \left.\frac{\partial \Omega}{\partial G'}\right|_{m, \mu_q} \frac{\partial G'(\mu_q)}{\partial \mu_q}.
\end{equation}
Since the physical mass is determined by the stationary condition $\partial \Omega / \partial m = 0$, the second term vanishes. Evaluating the remaining partial derivatives, the density takes the form
\begin{equation}
n_q = -\left.\frac{\partial\Omega_{\rm med}}{\partial\mu_q}\right|_{T,m} + \frac{(m-m_0)^2}{4 [G'(\mu_q)]^2} \frac{\partial G'(\mu_q)}{\partial\mu_q},
\label{eq:quark_density}
\end{equation}
where the second term guarantees thermodynamic stability and causal consistency when solving for the equation of state. The physical baryon number density $n_B$ and baryon chemical potential $\mu_B$ are related to the quark variables by
\begin{equation}
n_B = \frac{1}{3}n_q, \qquad \mu_B = 3\mu_q.
\label{eq:baryon_vars}
\end{equation}

The entropy density $s$ and pressure $p$ follow the standard definitions
\begin{eqnarray}
s &{=}& -\left.\frac{\partial\Omega}{\partial T}\right|_{\mu_q}, \label{eq:entropy} \\
p &{=}& -\Omega + \Omega_0, \label{eq:pressure}
\end{eqnarray}
where $\Omega_0 = \Omega(0,0,m_{\rm vac})$ is the vacuum subtraction that ensures $p=0$ at $T=\mu_q=0$. 

The speed of sound at constant entropy is defined as
\begin{equation}
c_s^2 = \left.\frac{\partial p}{\partial \varepsilon}\right|_{s},
\label{eq:cs2_def}
\end{equation}
where $\varepsilon = -p + Ts + \mu_q n_q$ is the energy density. This can be written in terms of susceptibilities
\begin{equation}
c_s^2 = \frac{n_q\chi_{TT} - s\chi_{\mu_q T}}{\mu_q\,(\chi_{TT}\chi_{\mu_q\mu_q}-\chi_{\mu_q T}^2)}\,,
\label{eq:cs2_full}
\end{equation}
where $\chi_{xy} = -\partial^2\Omega/\partial x\,\partial y$. In the low-temperature limit $T\to 0$, the entropy density vanishes ($s\to 0$ and $\chi_{\mu_q T}\to 0$), and the expression reduces purely to the density response
\begin{equation}
c_s^2\big|_{T\to 0} \approx \frac{n_q}{\mu_q}\left(\frac{\partial n_q}{\partial\mu_q}\right)^{-1} = \frac{n_B}{\mu_B}\left(\frac{\partial n_B}{\partial\mu_B}\right)^{-1},
\label{eq:cs2_T0}
\end{equation}
which is the form used in the numerical analysis of Sec.~\ref{sec:results}. This approximation is valid when $T \ll \mu_q$ and $T \ll T_c$; it will overestimate $c_s^2$ when thermal contributions to the susceptibilities become non-negligible near the transition boundaries.

\section{Numerical Results}\label{sec:results}

\begin{figure}[htbp]
\centering
\includegraphics[width=\textwidth]{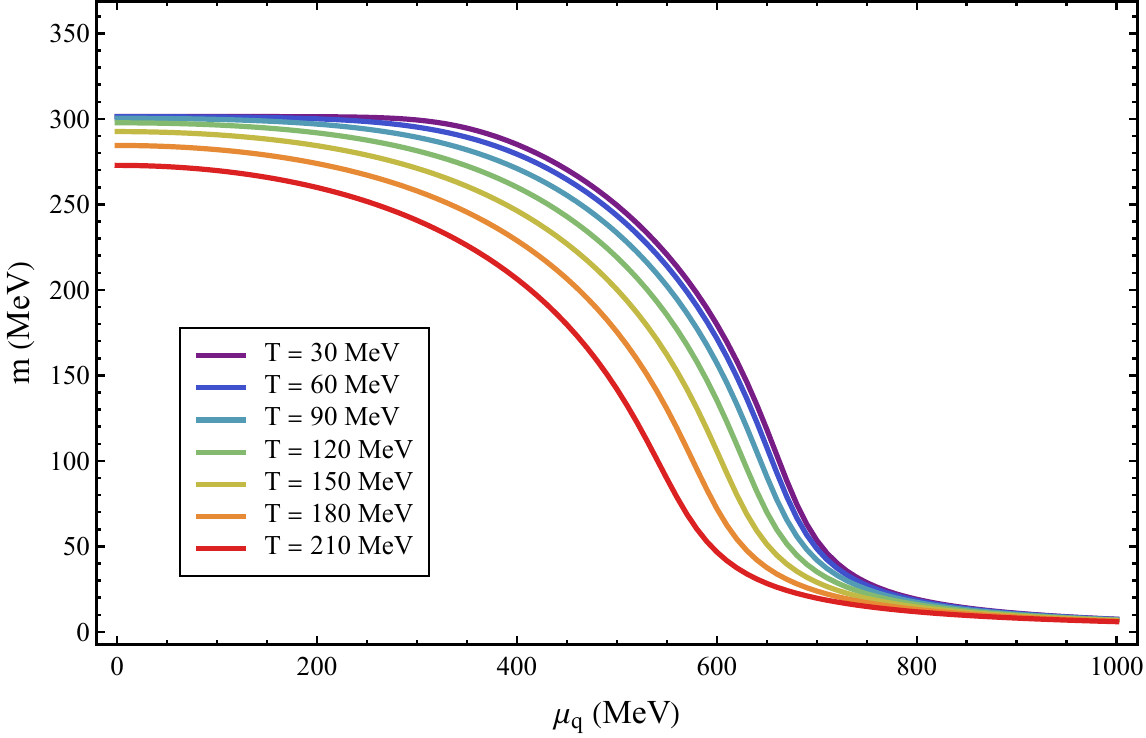}
\caption{Dynamical quark mass $m$ as a function of the quark chemical potential $\mu_q$ using a constant coupling $G$. The curves correspond to seven different temperatures, ranging from $T = 30$\,MeV to $210$\,MeV in increments of $30$\,MeV. At $\mu_q = 0$, the mass begins at the vacuum constituent value $m_{\text{vac}} \approx 301.38$\,MeV for low temperatures, reflecting the spontaneous breaking of chiral symmetry. As the chemical potential increases, chiral symmetry is progressively restored, and the dynamical mass drops asymptotically toward the current quark mass $m_0 = 5$\,MeV. At low temperatures, the transition is remarkably sharp, signaling the proximity of a first-order phase transition at $T \to 0$, whereas at higher temperatures, thermal fluctuations smooth the transition into a crossover.}
\label{fig:mass_constant_G}
\end{figure}

Throughout this section, we use the calibrated parameter set listed in Table~\ref{tab:model_parameters}: $m_0 = 5$\,MeV, $\Lambda = 1133.56$\,MeV, and $G\Lambda^2 = 2.428$, fixed to reproduce the vacuum pion mass $m_\pi = 138$\,MeV, the pion decay constant $f_\pi = 93$\,MeV, and the quark condensate $\langle\bar uu\rangle^{1/3} = -250$\,MeV. Before investigating the thermodynamic implications of the medium-dependent screening coupling $G'(\mu_q)$, we establish the baseline behavior of the chiral transition using a constant coupling $G$. We solve the modified analytical gap equation across the full phase diagram using in-medium Fermi-Dirac integrals, tracking the physical branch that identifies spontaneous chiral symmetry breaking in the vacuum. Figure~\ref{fig:mass_constant_G} illustrates these baseline solutions for several temperatures.

\begin{figure}[htbp]
    \centering
    \includegraphics[width=1\textwidth]{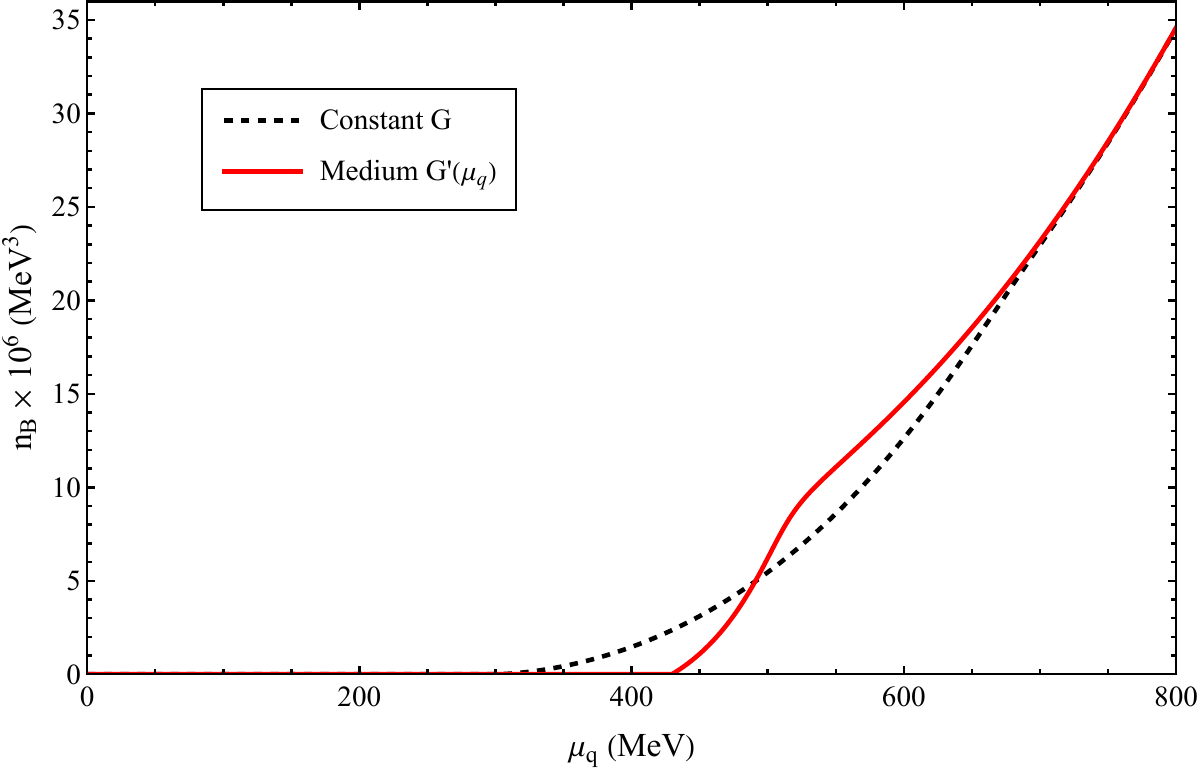}
    \caption{Baryon number density $n_B(\mu_q)$ as a function of the quark chemical potential in the cold degenerate regime at $T = 10^{-5}$\,MeV, computed from $n_B = -\frac{1}{3} d\Omega/d\mu_q$. The dashed black curve corresponds to the baseline model with constant coupling $G$, and the solid red curve incorporates the screening coupling $G'(\mu_q)$ together with its Hugenholtz--Van Hove rearrangement term. The negative contribution of the rearrangement term ($\partial G'/\partial \mu_q < 0$) acts as an effective repulsive mean-field penalty, delaying the physical onset of quark matter from $\mu_q \approx 301.38$\,MeV up to $\mu_q \approx 430$\,MeV and stiffening the equation of state in the transitional region.}
    \label{fig:density_comparison}
\end{figure}

Because the dynamical mass is directly linked to the chiral condensate $\langle\bar\psi\psi\rangle$, changes in $m(\mu_q)$ translate directly into modifications of macroscopic thermodynamic observables. For a constant coupling $G$, the baryon number density is computed directly from the relation $n_B = \frac{1}{3} n_q = -\frac{1}{3} \partial\Omega/\partial\mu_q$. Figure~\ref{fig:density_comparison} shows this baseline density profile in the cold degenerate limit ($T = 10^{-5}$\,MeV, dashed black curve), where $n_B$ remains zero until the chemical potential overcomes the vacuum constituent mass gap ($\mu_q \approx m_{\text{vac}} \approx 301.38$\,MeV), marking the physical onset of quark matter.

\begin{figure}[htbp]
    \centering
    \includegraphics[width=\textwidth]{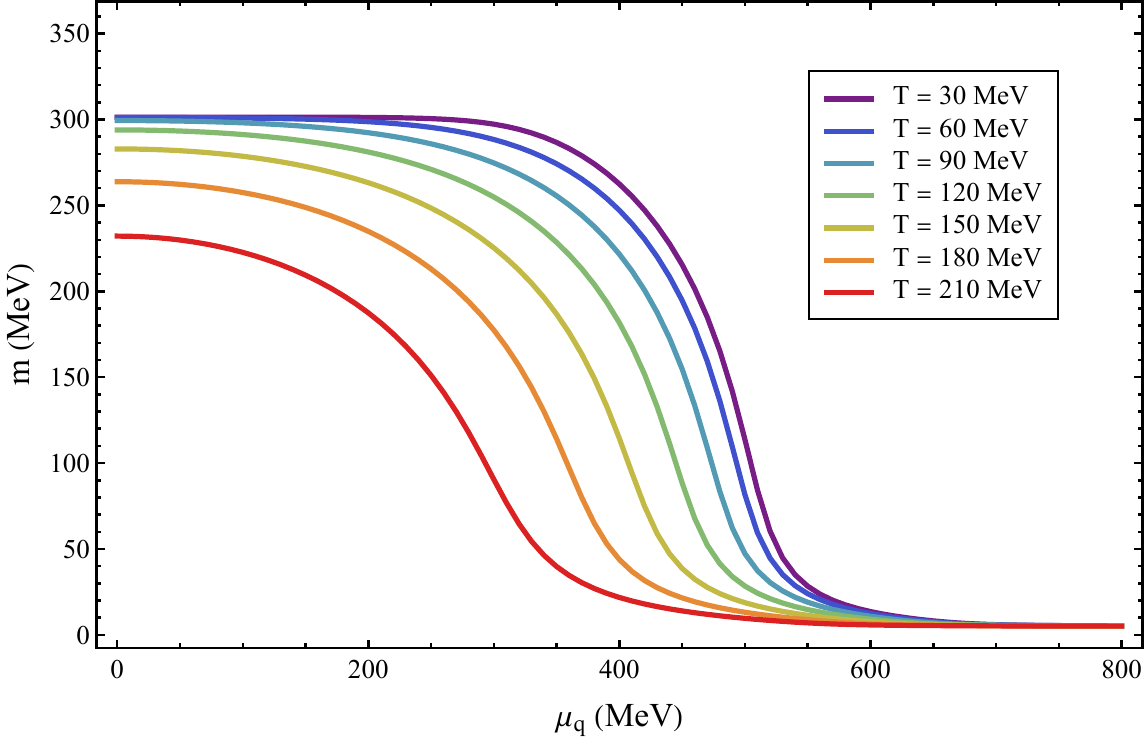}
    \caption{Dynamical quark mass $m$ as a function of the quark chemical potential $\mu_q$, incorporating the screening coupling $G'(\mu_q) = G(m/m_{\text{vac}})$. The curves represent seven distinct temperatures ranging from $T = 30$\,MeV to $210$\,MeV in increments of $30$\,MeV. Compared to the constant coupling (Fig.~\ref{fig:mass_constant_G}), the scaling of $G'$ introduces a pronounced positive feedback mechanism: as the rising density depletes the chiral condensate and lowers the dynamical mass, the effective scalar interaction strength $G'$ weakens precipitously. This thermodynamic suppression accelerates the restoration of chiral symmetry, causing the constituent mass to collapse toward the current quark mass ($m_0 = 5$\,MeV) at lower chemical potentials ($\mu_q \approx 500$\,MeV) and with sharper phase boundaries.}
    \label{fig:mass_running_G_prime}
\end{figure}

Having established the baseline properties of the chiral phase transition and the baryon density with a constant coupling, we now investigate the thermodynamic implications of the screening coupling $G'(\mu_q) = G(m/m_{\text{vac}})$. As discussed in Sec.~\ref{sec:framework}, to maintain the form of the thermodynamic potential without introducing unphysical non-linear order-parameter derivatives into the gap equation, the modified coupling is treated as an explicit function of the medium properties. Specifically, $G'(\mu_q)$ is evaluated along the physical trajectory using the baseline constituent mass derived from the constant $G$ scenario. The resulting dynamical quark mass under this screened interaction is presented in Figure~\ref{fig:mass_running_G_prime}.

With the dynamical mass established under the screened coupling, we now compute the corresponding baryon number density. Crucially, as derived in Eq.~\eqref{eq:quark_density}, the evaluation of the density must incorporate the thermodynamic rearrangement term to ensure stability, causality, and compliance with the Hugenholtz--Van Hove identity. The resulting density profile at $T = 10^{-5}$\,MeV (solid red curve) is compared against the constant-coupling baseline in Figure~\ref{fig:density_comparison}.

The delayed onset observed in Fig.~\ref{fig:density_comparison} highlights the impact of the rearrangement term. By penalizing the creation of baryon density due to the energetic cost of modifying the background interaction strength, the medium-dependent scalar coupling naturally generates a repulsive-like vector interaction effect. This stiffening of the dense fluid is a key mechanism for supporting high-mass compact stars and has immediate consequences for the speed of sound.

\begin{figure}[htbp]
    \centering
    \includegraphics[width=0.8\textwidth]{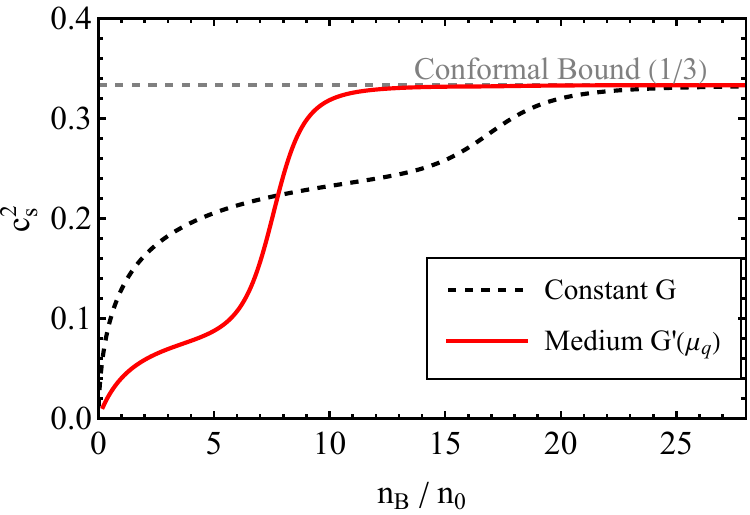} 
    \caption{Squared speed of sound $c_s^2$ as a function of the normalized baryon number density $n_B / n_0$ at $T=0$, where $n_0 = 0.16\,\text{fm}^{-3}$ ($\approx 1.23 \times 10^6\,\text{MeV}^3$) denotes the nuclear saturation density. The black dashed curve corresponds to the baseline NJL model with a constant coupling $G$, while the solid red curve incorporates the screening coupling $G'(\mu_q)$ and its thermodynamic rearrangement term. The horizontal gray dashed line marks the ultrarelativistic conformal bound ($c_s^2 = 1/3$). Expressed in units of nuclear saturation density, the effective repulsion generated by the rearrangement term induces a rapid stiffening of the equation of state at $n_B / n_0 \approx 6\text{--}10$, reaching the asymptotic conformal limit earlier than the constant coupling scenario while preserving causality ($c_s^2 \leq 1$).}
    \label{fig:cs2_vs_density}
\end{figure}

To examine the astrophysical implications of the stiffened equation of state, we evaluate the squared speed of sound at zero temperature using the density response relation given in Eq.~\eqref{eq:cs2_T0}
\begin{equation}
    c_s^2 = \frac{n_B}{\mu_B} \left( \frac{\partial n_B}{\partial \mu_B} \right)^{-1} = \frac{n_B}{\mu_q} \left( \frac{\partial n_B}{\partial \mu_q} \right)^{-1}.
\end{equation}
While the quark chemical potential $\mu_q$ is the natural control variable of the grand canonical ensemble, the response of the dense fluid with respect to its physical baryon number density $n_B$ governs the macroscopic hydrostatics and stability of compact stars. By parametrically mapping $c_s^2(\mu_q)$ against the corresponding normalized baryon density $n_B(\mu_q)/n_0$, we compare the speed of sound for both the constant coupling baseline and the medium-modified framework in Figure~\ref{fig:cs2_vs_density}.

At the onset of quark matter ($n_B / n_0 \to 0$), the quark gas is highly dilute and compressible in both models, resulting in $c_s^2 \to 0$. In the constant coupling scenario (black dashed curve), the equation of state remains relatively soft, stiffening only gradually over a wide density range; its subtle inflection at intermediate densities ($n_B / n_0 \approx 8\text{--}16$) is a signature of the progressive restoration of chiral symmetry before the curve smoothly approaches the ideal gas limit from below~\cite{Fraga2001,Kurkela2010}. Conversely, the introduction of the screened coupling $G'(\mu_q)$, counterbalanced by the Hugenholtz--Van Hove rearrangement term (solid red curve), produces a sharp stiffening at much lower baryon densities ($n_B / n_0 \approx 6\text{--}10$, corresponding to $n_B \approx 7\text{--}12 \times 10^6\,\text{MeV}^3$), where $n_0 = 0.16\,\text{fm}^{-3}$ is the nuclear saturation density. Despite this rapid rise in internal pressure, the rearrangement term ensures that the system respects the asymptotic limits of QCD, smoothly converging to the ultrarelativistic conformal bound ($c_s^2 \to 1/3$) at high densities without unphysical overshoots.

\section{Discussion and Perspectives}
\label{sec:discussion}

The results presented in Sec.~\ref{sec:results} illustrate how medium-dependent screening affects chiral symmetry restoration and the equation of state within an analytically regularized NJL framework. With a constant coupling $G$, the dynamical quark mass decreases monotonically as the quark chemical potential $\mu_q$ increases, with the onset of quark matter occurring at the vacuum constituent-quark threshold ($m_{\rm vac} \approx 301.38$\,MeV). However, when the dynamically screened coupling $G'(\mu_q)$ is introduced, the effective scalar interaction weakens at high densities. This introduces a positive feedback mechanism, analogous to the Debye screening of gluon exchanges in dense QCD~\cite{Roberts1994,Fischer2009,Farias2016,Farias2017}, which accelerates the destruction of the chiral condensate. Consequently, the structural changes in the chiral order parameter become sharper and shift to lower chemical potentials ($\mu_q \approx 500$\,MeV) compared to the constant coupling scenario.

The ansatz $G'(\mu_q) = G\,m(\mu_q)/m_{\rm vac}$ adopted here provides a minimal phenomenological approach to capture this in-medium screening effect. While running couplings derived from the gluon Debye mass~\cite{Schaefer2007} or Polyakov-loop-dependent interactions~\cite{Fukushima2004,Ratti2006} are also employed in the literature, our parametrization captures the physics of an in-medium weakening interaction without introducing additional free parameters. A more systematic microscopic derivation using Dyson-Schwinger equations for the in-medium gluon propagator~\cite{Fischer2009,Roberts2012} or functional renormalization group (FRG) methods~\cite{Tripolt2014,Braun2012} remains a natural extension of the present work.

Because the screened coupling runs with the chemical potential, the Hugenholtz--Van Hove identity dictates the inclusion of a thermodynamic rearrangement term. As demonstrated in our baryon density calculations (Fig.~\ref{fig:density_comparison}), this contribution ($\partial G'/\partial \mu_q < 0$) acts as an effective repulsive mean-field penalty. Even though the dynamical mass collapses at lower chemical potentials, this repulsion delays the physical onset of quark matter, pushing the threshold from $\mu_q \approx 301.38$\,MeV up to $\mu_q \approx 430$\,MeV.

This mechanism impacts the stiffness of the equation of state, which is most sensitively probed by the squared speed of sound $c_s^2$ mapped as a function of the normalized baryon number density $n_B/ n_0$ (Fig.~\ref{fig:cs2_vs_density}). For a constant coupling $G$, $c_s^2(n_B/n_0)$ rises smoothly and gradually toward the conformal bound from below, consistent with standard perturbative QCD expectations~\cite{Fraga2001,Kurkela2010}. Conversely, the effective repulsion generated by the rearrangement term for $G'(\mu_q)$ forces the dense fluid to stiffen rapidly after the delayed onset. The speed of sound rises steeply at intermediate baryon densities ($n_B / n_0 \approx 6\text{--}10$), yet the system respects causality ($c_s^2 \leq 1$) and smoothly asymptotes to the ultrarelativistic conformal limit ($c_s^2 \to 1/3$) at high densities.

This rapid stiffening of the equation of state at intermediate to high densities is relevant to the physics of compact stars, where core densities may exceed several times the nuclear saturation density~\cite{Ozel2016,Annala2020}. The ability of the medium-modified framework to dynamically generate a stiff equation of state, effectively mimicking the repulsive behavior typically obtained by introducing explicit vector-channel interactions~\cite{Buballa2005,Costa2008,Bratovic2013}, while preserving the asymptotic conformal limit places this model within the regime of phenomenological interest for supporting massive neutron stars~\cite{Demorest2010,Antoniadis2013,Fonseca2021,Abbott2018}. Evaluating the macroscopic mass-radius relation by integrating the Tolman--Oppenheimer--Volkoff (TOV) equations, as well as incorporating strange quarks, diquark condensates, and a proper hadronic-quark matter interface~\cite{Alford2005,Baym2018,Annala2020}, constitutes a primary direction for subsequent astrophysical applications.

Several approximations adopted in this study should be noted. The Hartree approximation omits mesonic fluctuations that can quantitatively modify the transition temperatures and chemical potentials~\cite{Schaefer2007,Herbst2011}; Within that approximation, the shift in the transition chemical potential reported in Fig.~\ref{fig:density_comparison} is a mean-field estimate, mesonic fluctuations are expected to move it by at most a few tens of MeV, small against the $\approx 130\,$MeV delay the rearrangement term produces. The isospin-symmetric limit does not account for $\beta$-equilibrium conditions relevant to stellar matter; at the delayed threshold of $\mu_q \approx 430\,$MeV, the splitting between $u$ and $d$ chemical potentials set by $\beta$-equilibrium is of order $10\text{-}20\,$MeV, small compared to the same $\approx 130\,$MeV delay, so the isospin-symmetric result stands as the leading estimate.

Future work will address these extensions by solving the hydrostatic TOV equations to quantify the maximum stellar mass supported by the rearrangement-induced stiffening, incorporating Polyakov-loop degrees of freedom~\cite{Fukushima2004,Ratti2006}, comparing the emergent repulsion against explicit vector channels~\cite{Costa2008,Bratovic2013}, and extending the formalism to the three-flavor sector~\cite{Alford2008}. Computing the quark-number susceptibilities and baryon-number fluctuations across the full finite-temperature phase diagram will also enable direct comparisons with lattice QCD data~\cite{Borsanyi2012,Bazavov2020} and freeze-out analyses of heavy-ion collisions~\cite{Adamczyk2014,Luo2017}. The modified regularization scheme established in this work provides a computationally stable and thermodynamically consistent foundation for these future investigations.

\section*{Acknowledgements}
A.\ Raya and S.H.O. acknowledge support from SECIHTI and CIC-UMSNH (Mexico) under grants
CBF-2025-G-1718 and 18371, respectively. S.H.O. acknowledges support from a SECIHTI (Mexico)
postdoctoral project under grant EPA(3)-2022. C.A.V.-A. is supported by the Secretar\'ia de Ciencia, Humanidades, Tecnolog\'ia e Innovaci\'on (Secihti) Investigadoras e Investigadores por M\'exico project 749 and SNII 58928

\bibliography{referencias}
\bibliographystyle{ieeetr}

\end{document}